\documentclass[conference,letterpaper]{IEEEtran}
\usepackage[T1]{fontenc}
\usepackage{amsmath,amssymb}
\usepackage{graphicx}
\usepackage{booktabs}
\usepackage{cite}
\usepackage{microtype}
\PassOptionsToPackage{hyphens}{url}
\usepackage[hidelinks]{hyperref}
\graphicspath{{figures/}}
\hypersetup{
    pdftitle={AC Assessment of DC Power Flow Screening for Data Center Hosting Capacity and Siting},
    pdfauthor={Haoxiang Wan, Jesus Silva-Rodriguez, Jonghwan Kwon, and Xingpeng Li}
}
\newcommand{\jj}{\mathrm{j}}
\newcommand{\ucom}{\hat u^{b}_{gt}}
\DeclareMathOperator*{\argmax}{arg\,max}

\title{AC Assessment of DC Power Flow Screening for Data Center Hosting Capacity and Siting}
\author{\IEEEauthorblockN{Haoxiang Wan\textsuperscript{1}, Jesus Silva-Rodriguez\textsuperscript{2}, Jonghwan Kwon\textsuperscript{3}, and Xingpeng Li\textsuperscript{1}}
\IEEEauthorblockA{\textsuperscript{1}Department of Electrical and Computer Engineering, University of Houston, Houston, TX, USA}
\IEEEauthorblockA{\textsuperscript{2}Electric Reliability Council of Texas, Taylor, TX, USA}
\IEEEauthorblockA{\textsuperscript{3}Energy Systems and Infrastructure Analysis Division, Argonne National Laboratory, Lemont, IL, USA}}

\begin{document}
\maketitle

\begin{abstract}
A screening procedure based on direct current (DC) power flow estimates
the data center hosting capacity of a candidate bus, termed the
DC-screened capacity, and selects the generator commitment that supports
it. This paper assesses that estimate under the alternating current (AC)
network model. With the selected commitment fixed, an AC optimal power
flow reoptimizes dispatch and voltages to find the largest data center
load the bus can carry, termed the AC-assessed capacity. On a synthetic
Texas system with 123 buses, the DC-screened capacity exceeds the
AC-assessed capacity at 113 buses. Across all 123, the ratio of
DC-screened to AC-assessed capacity has a median of 143\% (a 43\%
overestimate) and a maximum of 1,070\%. At these 113 buses, the
DC-screened capacity violates a branch apparent power limit or a
generator reactive power limit in the AC model, two limits the DC model
omits. The overestimate changes decisions: a 1-GW project passes DC
screening at every bus but exceeds the AC-assessed capacity at 37, and
the bus ranked first by DC screening has 205.5~MW less AC-assessed
capacity than the bus ranked second. Reporting the AC-assessed capacity
beside each DC-screened capacity lets planners reject oversized projects
or choose better sites.
\end{abstract}


\section{Introduction}\label{sec:intro}

Data centers consumed 4\% to 5\% of United States electricity in 2024
and are projected to consume 9\% to 17\% by 2030~\cite{epri2026}. The
large load interconnection queue of the Electric Reliability Council of
Texas (ERCOT) held requests totaling approximately 410~GW as of
March~2026, about 87\% of them from data centers~\cite{ercot2026}.
Screening a queue of that size means solving unit commitment and
expansion planning models at scale, and those models rely on the direct
current (DC) power flow approximation, which is linear in the voltage
angles and scales to thousands of buses and hours~\cite{stott2009}. The
approximation gains that tractability by dropping losses, holding
voltage magnitudes fixed, and omitting reactive power~\cite{purchala2005};
a dispatch that is optimal under it need not be feasible under the
alternating current (AC) network model~\cite{baker2021}.

For data centers, these omitted quantities matter. Emerging large loads
include complexes of several gigawatts connected at the transmission
level; in data centers, the power electronic drives of the cooling
systems can increase reactive power demand, and large loads in general
can affect voltage response and stability~\cite{nerc2025large}. A 2026 alert from
the North American Electric Reliability Corporation (NERC) calls for the
expected power factor of each computational load and for system studies
that may include active power versus voltage (P-V) and active versus
reactive power (P-Q) analysis~\cite{nerc2026alert}. These studies rest on reactive power and voltage, the two quantities the
DC approximation excludes.

Many studies of data center integration at the transmission level are
built on the DC model. Under it, spatial load flexibility has been valued for
congestion relief and curtailment reduction~\cite{wan2026spatial},
spatial and temporal flexibility have been coordinated within
security-constrained unit commitment~\cite{wan2026scuc}, and the
sensitivity of hyperscale hosting capacity to security modeling
fidelity has been examined~\cite{wan2026fidelity}. Two related lines
of work point at what such DC studies leave out: system-level analyses
show that large loads interconnect faster when they can be
curtailed~\cite{norris2025}, whereas a data center sited as a firm
load cannot rely on that relief; and at the distribution level, an AC branch flow model finds edge
data center hosting capacity bounded by branch current and voltage
limits that vary strongly with location~\cite{fang2026edge}. For a
candidate bus, DC screening estimates the hosting capacity, the largest
constant data center load the bus can accept, and selects the generator
commitment that supports it.

Two established approaches narrow the gap between the DC and AC network
models. A data-driven enhanced linearized AC model keeps voltage
magnitude and reactive power in a linear formulation and reduces its
error through regression~\cite{li2020}. Alternatively, the unit
commitment problem can include AC optimal power flow (ACOPF) constraints
directly~\cite{castillo2016}. Both approaches improve how a
plan is chosen. This paper takes the plan as given and assesses it under
the AC network model. Holding the generator unit commitment selected by
DC screening fixed, the AC assessment determines how much firm, constant
data center load a transmission bus can carry within generator reactive
power, bus voltage magnitude, and branch apparent power limits.

For each candidate bus $b$, let $H_b^{\mathrm{DC}}$ denote the
DC-screened capacity and $H_b^{\mathrm{AC}}$ the AC-assessed capacity
under the same commitment. Comparing the two shows how far the
DC-screened capacity departs from what the AC network can carry, and
whether the screened project size and site remain valid. The
contributions are threefold.

\begin{enumerate}
\item An assessment procedure that fixes the generator unit commitment
selected by DC screening and reoptimizes dispatch and voltages with
ACOPF to find the largest data center load each candidate bus can carry
under the AC network model.

\item A diagnosis of the AC limits violated at the DC-screened
capacity, obtained with penalized slack variables. At every candidate
whose DC-screened capacity is not AC feasible, the diagnosis identifies
a branch apparent power limit or a generator reactive power limit,
neither of which is represented in the DC formulation.

\item An evaluation of the DC-screened plans at all 123 buses of the
synthetic Texas backbone transmission system
(TX-123BT)~\cite{lu2025tx123}, over the 24 hours of its peak day.
\end{enumerate}

\section{Assessing a DC-Screened Plan}\label{sec:audit}

\subsection{Plan and Load Definition}\label{sec:what}

Let $\mathcal N$ be the set of buses, $\mathcal L$ the set of branches,
and $\mathcal T$ the set of hourly periods in one day. All active and
reactive power variables and limits in the formulations below are
expressed in per unit on a common system base; capacity values are
converted to MW for reporting.

Let $h\ge0$ denote the added active demand of the data center. If the
data center is located at bus $b\in\mathcal N$, its constant active and
reactive demands at bus $i\in\mathcal N$ and hour $t\in\mathcal T$ are
\begin{equation}
 P^{\mathrm{c}}_{it}=\delta_{ib}h,\qquad
 Q^{\mathrm{c}}_{it}=\kappa\delta_{ib}h,\qquad
 \kappa=\tan(\arccos \mathrm{pf}),
 \label{eq:load}
\end{equation}
where the superscript c denotes the added data center load,
$\delta_{ib}$ equals one for $i=b$ and zero otherwise, and
$\mathrm{pf}$ is the lagging power factor of the data center.

\begin{figure}[!t]
\centering
\includegraphics[width=0.86\columnwidth]{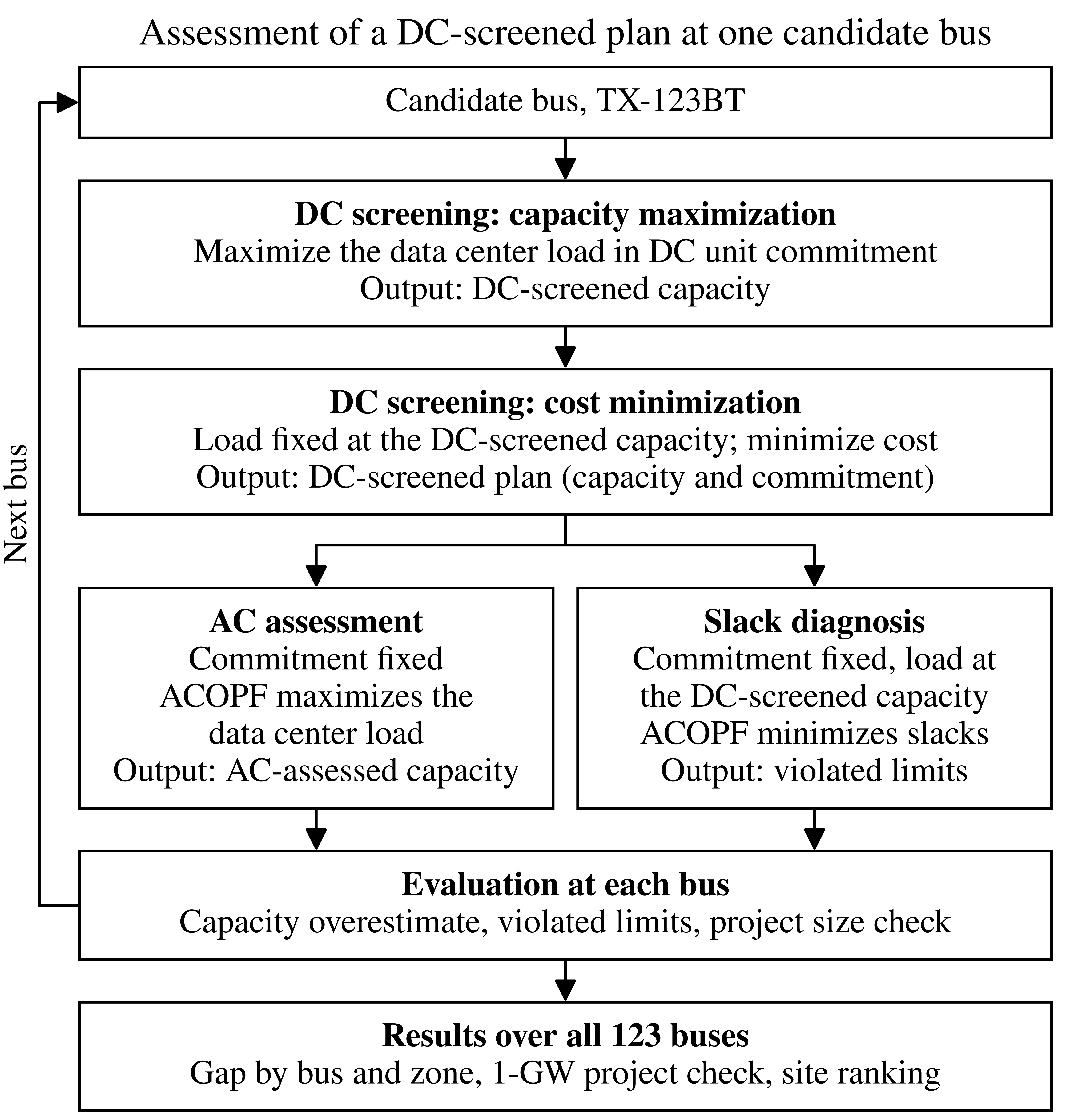}
\caption{Assessment procedure for a DC-screened plan, repeated at each of the 123 candidate buses.}
\label{fig:flow}
\end{figure}

Hosting capacity is the largest feasible $h$. DC screening maximizes
$h$ and returns the generator unit commitment that supports it; the AC
assessment fixes that commitment and maximizes $h$ again, now subject
to generator reactive power, bus voltage magnitude, and branch apparent
power limits.
Fig.~\ref{fig:flow} shows the complete procedure at one candidate bus.

DC screening maximizes $h$ in a unit commitment model. Its variables
include the active output $p_{gt}$ of each conventional or renewable
generator $g$, the commitment status $u_{gt}$ and reserve $r_{gt}$ of
each conventional generator, the bus voltage angle $\theta_{it}$, and
the active branch flow $f_{\ell t}$. Renewable output is bounded by its
hourly availability. Commitment transitions, minimum up and down times,
ramping, and generation limits are enforced using standard mixed integer
linear constraints~\cite{carrion2006,moralesespana2013}. In each hour,
the reserve held by the other units must cover the output of every online
unit, implementing the largest unit reserve rule. For every bus
$i\in\mathcal N$, branch $\ell=(m,n)\in\mathcal L$, and hour
$t\in\mathcal T$, the DC network constraints are
\begin{subequations}\label{eq:dc}
\begin{align}
 &\sum_{g\in\mathcal G_i} p_{gt}
  -(1+\lambda)\bigl(P^{\mathrm{d}}_{it}+P^{\mathrm{c}}_{it}\bigr)
  =\sum_{\ell\in\mathcal L} A_{i\ell}\, f_{\ell t}, \label{eq:dc_bal}\\
 &f_{\ell t} = \bigl(\theta_{mt}-\theta_{nt}\bigr)/x_\ell, \label{eq:dc_flow}\\
 &-F_{\ell t} \le f_{\ell t} \le F_{\ell t}. \label{eq:dc_lim}
\end{align}
\end{subequations}
Here, $\mathcal G_i$ is the set of conventional and renewable generators
at bus $i$, $P^{\mathrm{d}}_{it}$ is the native active demand,
$A_{i\ell}$ is the bus branch incidence matrix with outgoing orientation,
$x_\ell$ is the branch reactance, and $F_{\ell t}$ is the hourly active
flow limit. The factor $\lambda=0.02$ increases native and added active
demand by two percent to approximate losses omitted by the lossless DC
model.

The optimal value of $h$ is the DC-screened capacity $H_b^{\mathrm{DC}}$.
Because multiple commitments can attain the same $H_b^{\mathrm{DC}}$, a
second solve fixes $h=H_b^{\mathrm{DC}}$ and selects the commitment with
the lowest daily operating cost, including fuel, no-load, and startup
costs of the conventional units. The resulting commitment is denoted by
$\hat u^b$, with elements $\hat u^b_{gt}$. The pair
$(H_b^{\mathrm{DC}},\hat u^b)$ is the DC-screened plan for bus $b$.

\subsection{AC-Assessed Capacity Under the Selected Commitment}\label{sec:deliver}

The AC assessment fixes the generator commitment selected by DC
screening and solves an ACOPF problem that reoptimizes every continuous
variable. Let $V_{it}$ be the complex per unit voltage at bus $i$,
$Y_{ik}$ the entries of the per unit bus admittance matrix, $q_{gt}$ the
reactive output of generator $g$, $Q^{\mathrm{d}}_{it}$ the native
reactive demand, and $\jj$ the imaginary unit; an overbar denotes the
complex conjugate. For every bus
$i\in\mathcal N$ and hour $t\in\mathcal T$, the AC nodal balance is
\begin{equation}
\begin{split}
 &\sum_{g\in\mathcal G_i}(p_{gt}+\jj q_{gt})
 -(P^{\mathrm{d}}_{it}+P^{\mathrm{c}}_{it})
 -\jj(Q^{\mathrm{d}}_{it}+Q^{\mathrm{c}}_{it})\\
 &\hspace{25mm}
 =V_{it}\overline{\sum_{k\in\mathcal N}Y_{ik}V_{kt}}.
\end{split}
\label{eq:ac}
\end{equation}

Conventional generators remain subject to the selected commitment and
their active and reactive power limits, including
$\underline Q_g\ucom\le q_{gt}\le\overline Q_g\ucom$, where
$\underline Q_g$ and $\overline Q_g$ are the minimum and maximum
reactive output of generator $g$.
Renewable generation follows its hourly availability and reactive power
limits. Voltage magnitudes, branch angle differences, and apparent power
at both ends of each branch are bounded, while the ramping and reserve
requirements of DC screening are retained.

The AC assessment maximizes the added data center load $h$ subject to
these constraints. The optimal value of $h$ is the AC-assessed capacity
$H_b^{\mathrm{AC}}$, the largest load the selected commitment supports
under the AC network model.

\subsection{Where the DC-Screened Capacity Fails}\label{sec:diag}

An infeasible AC solve at $h=H_b^{\mathrm{DC}}$ does not reveal which
limit is violated. To find it, a second ACOPF problem fixes
$h=H_b^{\mathrm{DC}}$ under the same commitment and relaxes three limits
with nonnegative slacks: voltage magnitude, generator reactive power, and
branch apparent power. Each bus also receives an active power shortfall.
All other constraints remain exact, including the reactive nodal
balances, active generation limits, ramping limits, and reserve
requirements.

The problem minimizes the weighted sum of these slacks. If every slack is
zero, $H_b^{\mathrm{DC}}$ is AC feasible. Otherwise, the largest slack of
each type gives the violation of that limit and shows where the plan
fails.

\subsection{How the Gap Changes a Decision}\label{sec:decision}

The capacity overestimate of DC screening at bus $b$ is
\begin{equation}
 e_b=\frac{H_b^{\mathrm{DC}}-H_b^{\mathrm{AC}}}{H_b^{\mathrm{AC}}}\times 100\%,
 \label{eq:gap}
\end{equation}
the percentage by which the DC-screened capacity exceeds the AC-assessed
capacity. For example, $e_b=50\%$ means that the DC-screened capacity is
150\% of the AC-assessed capacity.

The two capacities also determine whether a project size is screened
correctly. A project of size $K$ at bus $b$ is a screening error if
\begin{equation}
H_b^{\mathrm{DC}} \ge K > H_b^{\mathrm{AC}},
\label{eq:threshold}
\end{equation}
that is, if it passes DC screening but exceeds the AC-assessed capacity.

The AC-assessed capacities also compare sites. Let
$a=\argmax_b H_b^{\mathrm{DC}}$ be the DC-preferred bus and $c$ any
other candidate. Both $H_a^{\mathrm{AC}}$ and $H_c^{\mathrm{AC}}$ are
assessed under their own DC-selected commitments, so a positive
$H_c^{\mathrm{AC}}-H_a^{\mathrm{AC}}$ means that the alternative can
carry more load than the bus that DC screening prefers.

\section{Test System and Evaluation Design}\label{sec:setup}

\begin{figure*}[!t]
\centering
\includegraphics[width=0.88\textwidth]{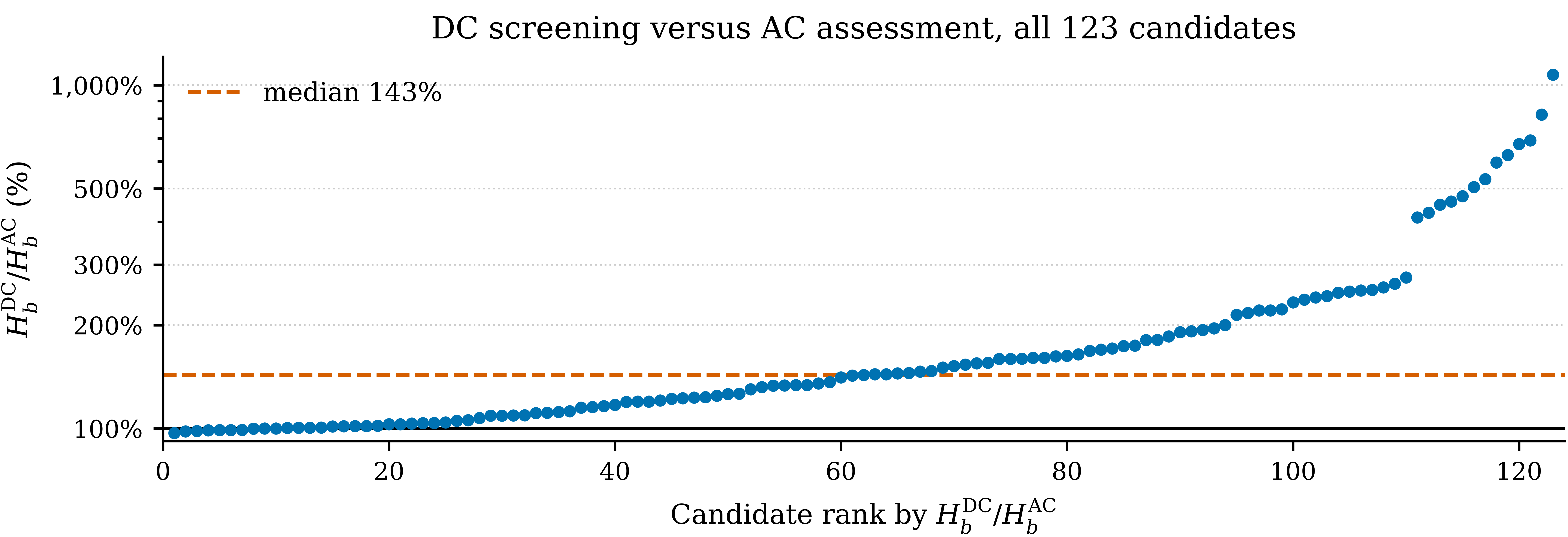}
\caption{DC-screened capacity as a percentage of the AC-assessed
capacity at each of the 123 candidates, in ascending order. Values above
100\%, found at 113 candidates, mean that DC screening overestimates the
capacity. The solid line marks 100\% and the dashed line the median,
143\%; the vertical axis is logarithmic.}
\label{fig:gaps}
\end{figure*}
The test system is the modified TX-123BT case~\cite{lu2025tx123}: 123
buses, 255 branches, 138 conventional and 154 renewable generators, with
the wind, solar, and hourly line rating profiles of its time series and an
ERCOT weather zone label on every bus. The eight zones are Coast (COAST),
East (EAST), Far West (FWEST), North (NORTH), North Central (NCENT),
South (SOUTH), South Central (SCENT), and West (WEST). All 123 buses are
evaluated over the 24 hours of the case's peak day.

\begin{table}[!t]
\caption{Common Conditions for All Candidate Assessments}
\label{tab:setup}
\centering\footnotesize
\begin{tabular}{@{}ll@{}}
\toprule
Quantity & Setting\\
\midrule
Data center load & Constant $P$, $Q$ at the candidate bus\\
Data center power factor & 0.98 lagging\\
Voltage magnitude bounds & 0.94 to 1.06 per unit\\
DC loss allowance & 2\% of native and added load\\
AC commitment & Candidate's DC-selected commitment\\
\bottomrule
\end{tabular}
\end{table}

Table~\ref{tab:setup} lists the conditions shared by all assessments.
Hourly line ratings apply as MW limits in DC and as numerically equal
MVA limits at both ends of each branch in AC.

Computations ran on the Carya cluster at the University of
Houston, using Intel Xeon Gold 5320 CPUs, with
candidates solved in parallel; the study took 59.5~h of aggregate
compute time, 7.2~h for DC screening of the 123 candidates and 52.3~h
for the AC assessments. Gurobi~13.0.2 solves both the capacity
maximization and cost minimization stages of DC screening using a
relative mixed integer optimality gap tolerance of $10^{-4}$, with
optimal status reported at every candidate bus. PowerModels.jl on
JuMP formulates the ACOPF problems, which
Ipopt solves. In the diagnosis problem, all slacks
are expressed in per unit on the 100~MVA system base, and the weighted
objective assigns a weight of 100 to the voltage magnitude slacks and
active power shortfalls and a weight of 1 to the generator reactive
power and branch apparent power slacks.

\section{Results: Capacity, Project Size, and Location}\label{sec:results}
\subsection{The Gap Between DC-Screened and AC-Assessed Capacity}
The DC-screened capacity exceeds the AC-assessed capacity at 113 of the
123 candidates. Over all 123, the capacity overestimate $e_b$ has a
median of 43.2\%, so at the median site the DC-screened capacity is
143\% of the AC-assessed capacity. Fig.~\ref{fig:gaps} shows the full
distribution, and it has two parts. Across 110 candidates,
$H_b^{\mathrm{DC}}/H_b^{\mathrm{AC}}$ rises without a break from 97\% to
275\%; the remaining 13 sit apart, between 410\% and 1,070\%, with
nothing in between. Brownsville~2 (bus~72) closes the sequence at
1,070\%, the largest value in the study.

At 10 candidates, $H_b^{\mathrm{DC}}/H_b^{\mathrm{AC}}$ is below 100\%
and $e_b$ is negative: the AC-assessed capacity exceeds the DC-screened
capacity. DC screening inflates every load by a fixed 2\% to cover
losses, and at these ten sites the actual AC losses are smaller than
that, so the AC model has slightly more room than DC screening allowed.

\subsection{What Breaks at the DC-Screened Capacity}
The AC model, unlike the DC model, includes the reactive demand of the
data center. By~\eqref{eq:load} with a 0.98 lagging power factor, a
1-GW data center adds 203.06~Mvar at its bus. The slack diagnosis fixes
$h$ at $H_b^{\mathrm{DC}}$ in the AC model, with the commitment
unchanged. Ten of the 123 plans pass this check with every slack at
zero.

The check fails at the other 113, and every one of the returned
diagnostic solutions violates a branch apparent power limit (111 plans),
a generator reactive production limit (100), or both; 29 also violate a
reactive absorption limit and 33 a voltage limit (11 high only, 11 low
only, 11 both). The violations are large. Among the 111 plans with a
branch apparent power violation, the largest violation of any branch in
any hour is 126~MVA at the median plan and 1,003~MVA at the most; among
the 100 with a reactive production violation, the largest excess of any
generator in any hour is 528~Mvar at the median plan and 2,396~Mvar at
the most, and
at Plano~2 (bus~114) it is 923~Mvar. The active power shortfall, by
contrast, stays below 0.001~MW at all 113 plans. Active generation
therefore covers the DC-screened capacity in full; the returned solution
exceeds limits because of the reactive power the load brings with it and
the apparent power headroom it consumes on the branches that carry it.
Neither limit exists in the DC model: it bounds active flow alone and
cannot see the reactive component that pushes apparent power over the
same rating, and it carries no generator reactive capability at all.

\subsection{The Discrepancy Depends Strongly on Location}
\begin{figure}[!t]
\centering
\includegraphics[width=0.9\columnwidth]{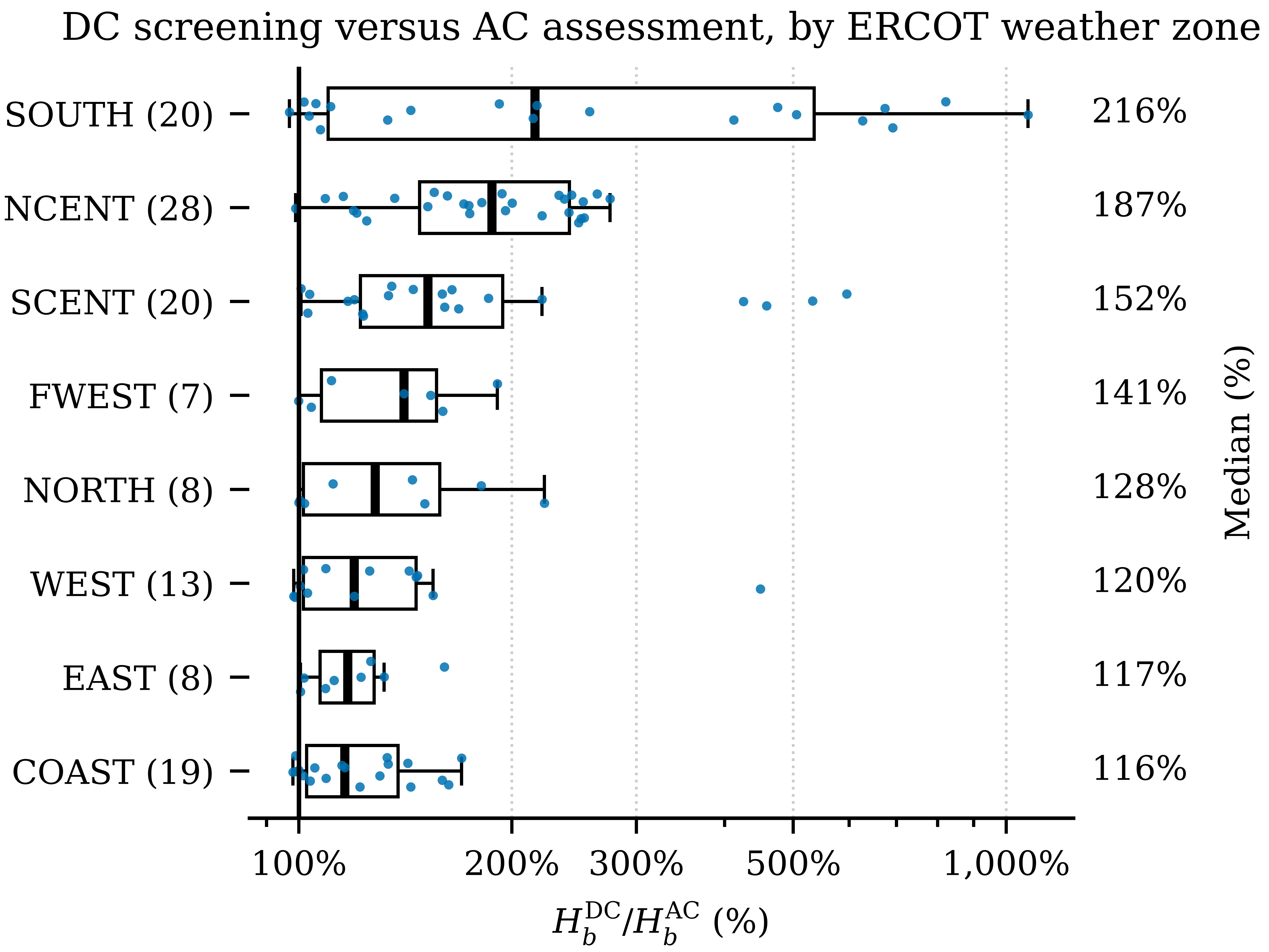}
\caption{The zonal median of $H_b^{\mathrm{DC}}/H_b^{\mathrm{AC}}$ ranges
from 116\% in COAST to 216\% in SOUTH. The horizontal axis is
logarithmic. Boxes span the interquartile range (IQR) with a median
line, whiskers extend to the most extreme candidates within 1.5 IQR of
the box, dots mark candidates, and parentheses give sample sizes.}
\label{fig:regions}
\end{figure}

Fig.~\ref{fig:regions} groups the 123 candidates by ERCOT weather zone.
The gap is heterogeneous in two ways. Between zones, the median of
$H_b^{\mathrm{DC}}/H_b^{\mathrm{AC}}$ moves from 116\% in COAST to 216\%
in SOUTH; within zones the spread is just as wide, WEST alone spanning
98\% to 449\%. The 13 candidates with $H_b^{\mathrm{DC}}/H_b^{\mathrm{AC}}$
above 400\% all lie in three zones: eight in SOUTH, four in SCENT, and
one in WEST. Four of the ten candidates whose AC-assessed capacity
exceeds the DC-screened capacity are in COAST. A single correction
factor, global or zonal, therefore cannot replace an AC assessment at
each bus.

\subsection{Project Sizes That Pass DC Screening but Exceed the AC-Assessed Capacity}\label{sec:reject}
Suppose a site is to be chosen for a 1-GW data center. All 123 candidate
buses have DC-screened capacities $H_b^{\mathrm{DC}}$ above 1~GW, so DC
screening passes every one of them. At 37 of them, however, the
AC-assessed capacity $H_b^{\mathrm{AC}}$ lies below 1~GW,
so~\eqref{eq:threshold} holds and each is a screening error. The 37 are
unevenly distributed: NCENT accounts for 20 and SOUTH for 8, while the
40 candidates of COAST, EAST, and WEST together have none.

\begin{figure}[!t]
\centering
\includegraphics[width=0.9\columnwidth]{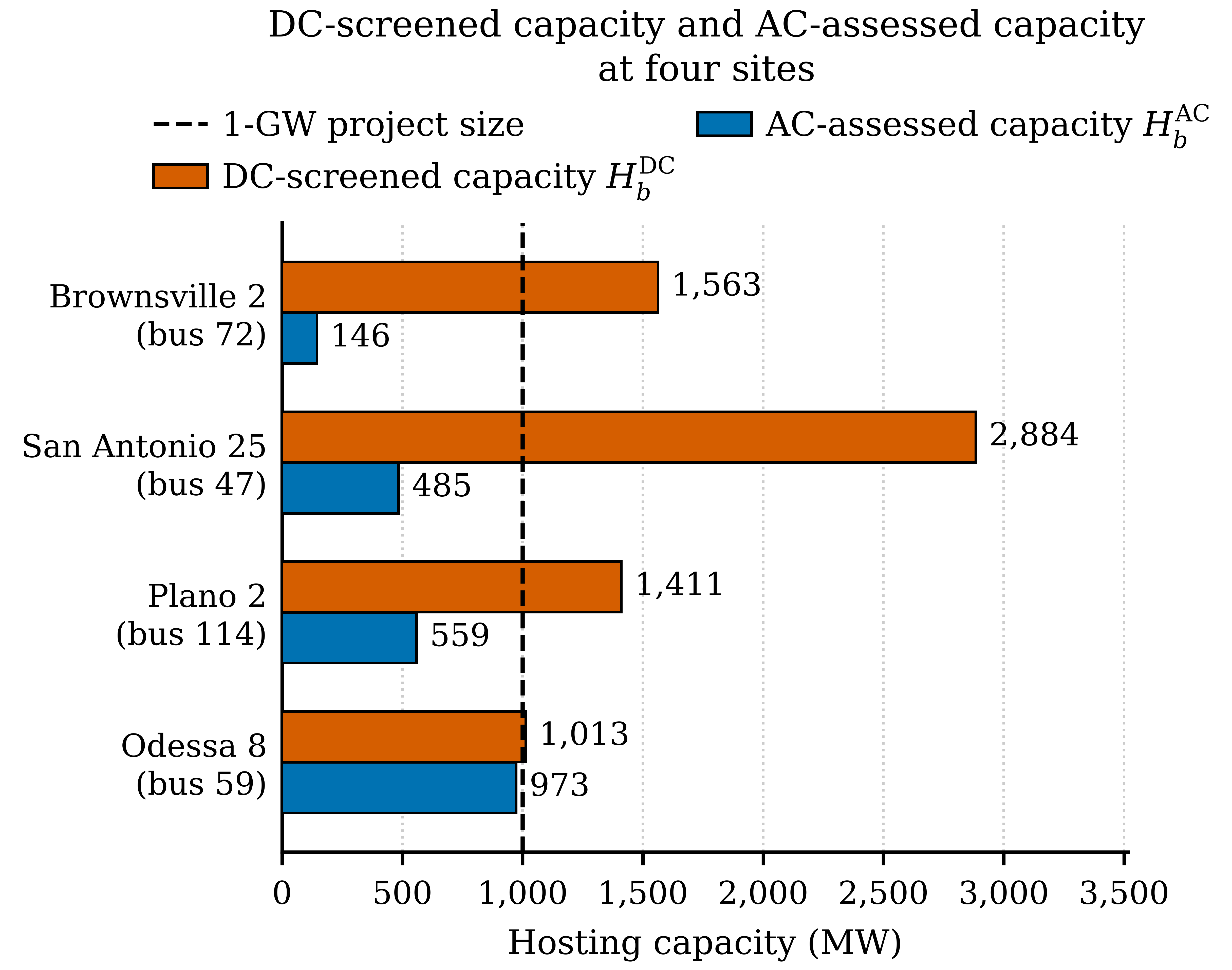}
\caption{Four of the 37 sites at which a 1-GW data center passes DC
screening but exceeds the AC-assessed capacity. Bars give the
DC-screened capacity $H_b^{\mathrm{DC}}$ and the AC-assessed capacity
$H_b^{\mathrm{AC}}$, labeled to the nearest MW. Each site is a separate
case under its own DC-selected commitment.}
\label{fig:promises}
\end{figure}

The margins differ widely (Fig.~\ref{fig:promises}). The AC-assessed
capacity at Brownsville~2 is 146~MW, 854~MW below 1~GW; at San
Antonio~25 (bus~47) it is 485~MW, 515~MW below; at Plano~2 it is 559~MW,
441~MW below; and at the closest case, Odessa~8 (bus~59), it is 973~MW,
27~MW below. There $H_{59}^{\mathrm{DC}}=1{,}013.01$~MW against
$H_{59}^{\mathrm{AC}}=973.28$~MW, so $e_{59}=4.08\%$: an overestimate of
four percent is enough to reverse a project decision.

\vspace{-1mm}\vspace{-1mm}
\subsection{The DC-Preferred Location Is Not the Best Plan}

\begin{figure}[!t]
\centering
\includegraphics[width=0.95\columnwidth]{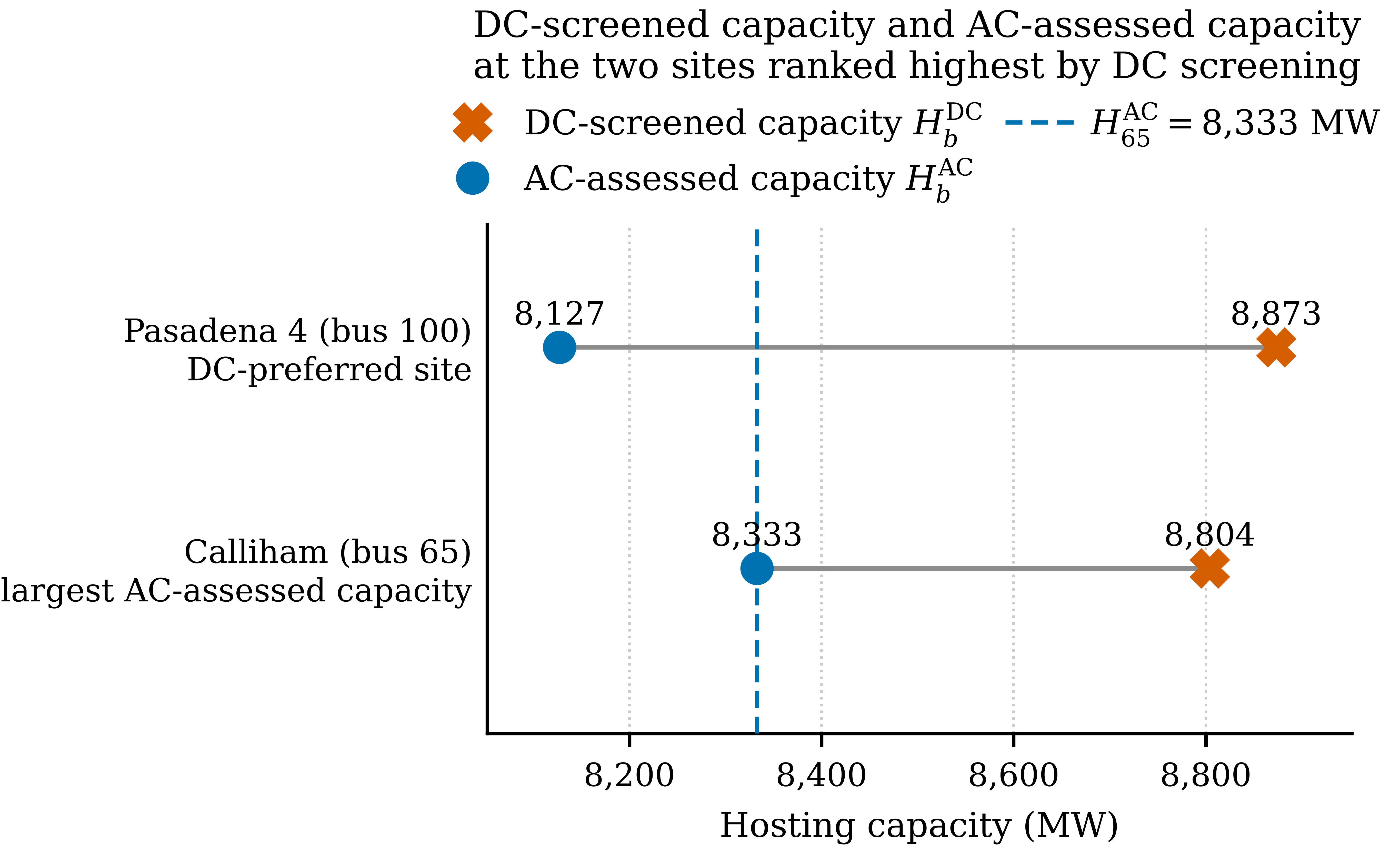}
\caption{DC screening prefers Pasadena~4 on $H_b^{\mathrm{DC}}$, while
Calliham has 205.5~MW more AC-assessed capacity. Circles give
$H_b^{\mathrm{AC}}$ and crosses $H_b^{\mathrm{DC}}$ at the two sites, and
the dashed line carries $H_{65}^{\mathrm{AC}}$ across both. The axis is
zoomed on the range that contains all four values.}
\label{fig:siting}
\end{figure}

DC screening prefers Pasadena~4 (bus~100) with
$H_{100}^{\mathrm{DC}}=8{,}873.25$~MW, the largest DC-screened capacity
in the study. Under that plan the AC-assessed capacity is
$H_{100}^{\mathrm{AC}}=8{,}127.36$~MW. Calliham (bus~65), which DC
screening ranks second with $H_{65}^{\mathrm{DC}}=8{,}804.11$~MW, has the
largest AC-assessed capacity of all 123 candidates,
$H_{65}^{\mathrm{AC}}=8{,}332.83$~MW (Fig.~\ref{fig:siting}), so
\begin{equation}
 H_{65}^{\mathrm{AC}}-H_{100}^{\mathrm{AC}}=205.5~\mathrm{MW}.
 \label{eq:numeric}
\end{equation}
Choosing the DC-preferred site therefore gives up 205.5~MW of
AC-assessed capacity relative to an alternative that DC screening ranks
lower.

\vspace{-1mm}
\vspace{-1mm}
\section{Conclusion}\label{sec:conclusion}

This paper assessed DC-screened data center hosting capacity under the
AC network model, holding fixed the generator commitment selected by DC
screening. On the peak day of the TX-123BT system, DC screening
overstates hosting capacity at 113 of the 123 candidate buses, and the
median overestimate across all 123 is 43\%. At each of these 113 buses,
active generation covers the DC-screened capacity, and the diagnostic
solution violates a branch apparent power limit or a generator reactive
power limit, neither of which the DC model represents. The overestimate
varies widely by location and changes decisions. A 1-GW project passes
DC screening at every bus but exceeds the AC-assessed capacity at 37 of
them; 28 of these are in NCENT and SOUTH, and none is among the 40
candidates of COAST, EAST, and WEST. The bus that DC screening ranks
first has 205.5~MW less AC-assessed capacity than the bus it ranks
second. Reporting the AC-assessed capacity beside each DC-screened
capacity gives planners a basis to reject a project size or revise a
site.

\vspace{-2mm}
\section*{Acknowledgment}
This work was supported in part by the University of Houston Energy Transition Institute Seed Grant Program and by computational resources from the Research Computing Data Core at the University of Houston.

\vspace{-1mm}

\bibliographystyle{IEEEtran}
\bibliography{references}
\end{document}